# Purcell radiative rate enhancement of label-free proteins with ultraviolet aluminum plasmonics

Aleksandr Barulin, Prithu Roy, Jean-Benoît Claude, Jérôme Wenger *

*Aix Marseille Univ, CNRS, Centrale Marseille, Institut Fresnel, 13013 Marseille, France*

* Corresponding author: jerome.wenger@fresnel.fr

**Abstract**

The vast majority of proteins are intrinsically fluorescent in the ultraviolet, thanks to the emission from their tryptophan and tyrosine amino-acid constituents. However, the protein autofluorescence quantum yields are generally very low due to the prevailing quenching mechanisms by other amino acids inside the protein. This motivates the interest to enhance the radiative emission rate of proteins using nanophotonic structures. Although there have been numerous reports of Purcell effect and local density of optical states (LDOS) control in the visible range using single dipole quantum emitters, the question remains open to apply these concepts in the UV on real proteins containing several tryptophan and tyrosine amino acids arranged in a highly complex manner. Here, we report the first complete characterization of the Purcell effect and radiative rate enhancement for the UV intrinsic fluorescence of label-free β-galactosidase and streptavidin proteins in plasmonic aluminum nanoapertures. We find an excellent agreement with a calibration performed using a high quantum yield UV fluorescent dye. Demonstrating and intensifying the Purcell effect is essential for the applications of UV plasmonics and the label-free detection of single proteins.

**Keywords:** Purcell effect, plasmonics, ultraviolet UV, autofluorescence, label-free protein, aluminum nanoapertures, zero-mode waveguide, single molecule

**Introduction**

The spontaneous emission depends on the photonic environment and is not an inherent process. This phenomenon, also called the Purcell effect [1], has been observed in photonic crystal microcavities [2,3] or plasmonic optical nanoantennas [4–8]. The Purcell effect finds its applications for single molecule microscopy [9], single photon sources [10,11], and sub-wavelength lasers [12].

While the spontaneous emission rate enhancement has been widely observed for single quantum emitters in the visible range, its extension into the UV domain remains essentially limited to simple fluorescent dyes such as p-terphenyl or isolated tryptophan amino acids [13–16]. Using the concepts of LDOS and Purcell effect with real proteins is complicated by the complex nature of the protein. Each system contains typically several tens of tryptophan emitters which are closely packed at a nanometer distance surrounded by other amino acids acting as fluorescence quenchers *via* electron or proton transfer [17,18]. The modifications of emission rates for specific proteins have been revealed in the optical range, including the pigment protein complex LH2 [19,20] and the green-fluorescent protein GFP [21]. However, these cases refer to specific pigment protein families featuring high absorption and emission in the visible, it cannot account for the collective UV emission of tryptophan and tyrosine amino acids present in the majority of proteins. Indeed, most proteins in nature contain those amino acids and, hence, can be potentially studied label-free in the ultraviolet range using conventional fluorescence microscopy. Boosting the radiative rate in the UV is of great importance as proteins exhibit inherently low quantum yield autofluorescence [18,22]. However, a clear demonstration of the Purcell enhancement of the intrinsic protein radiative rate in the UV has not been reported yet. The limited photon count rate in our previous work prevented the clear observation of any change in the fluorescence lifetime of label-free β-galactosidase protein [14].

Here, we report the first complete characterization of the Purcell effect and radiative rate enhancement for the UV intrinsic fluorescence of label-free β-galactosidase and streptavidin proteins in plasmonic aluminum nanoapertures (Fig. 1a-d). Aluminum (Al) is a well-established metal for UV plasmonics [23–27], and its applicability has already been reported for label-free sensing of biomolecules at high concentrations [28], as well as for UV photocatalysis [29], and optoelectronics [23]. We use a combination of measurements based on fluorescence correlation spectroscopy (FCS) and time-correlated single photon counting (TCSPC) to quantify the different photokinetic rates contributing to the fluorescence process [30]. Our approach allows us to completely map the different physical contributions which give rise to the observation of enhanced fluorescence emission for single label-free proteins inside a plasmonic nanoaperture. Our results stand in excellent agreement with a separate calibration using the high quantum yield UV fluorescent dye p-quaterphenyl. This importantly shows ways to improve the net detected autofluorescence signal by increasing the radiative emission and the autofluorescence quantum yield. The clear demonstration of Purcell effect applied to real label-free proteins is relevant for the emerging field of UV plasmonics and the applications of single molecule microscopy [23,31].

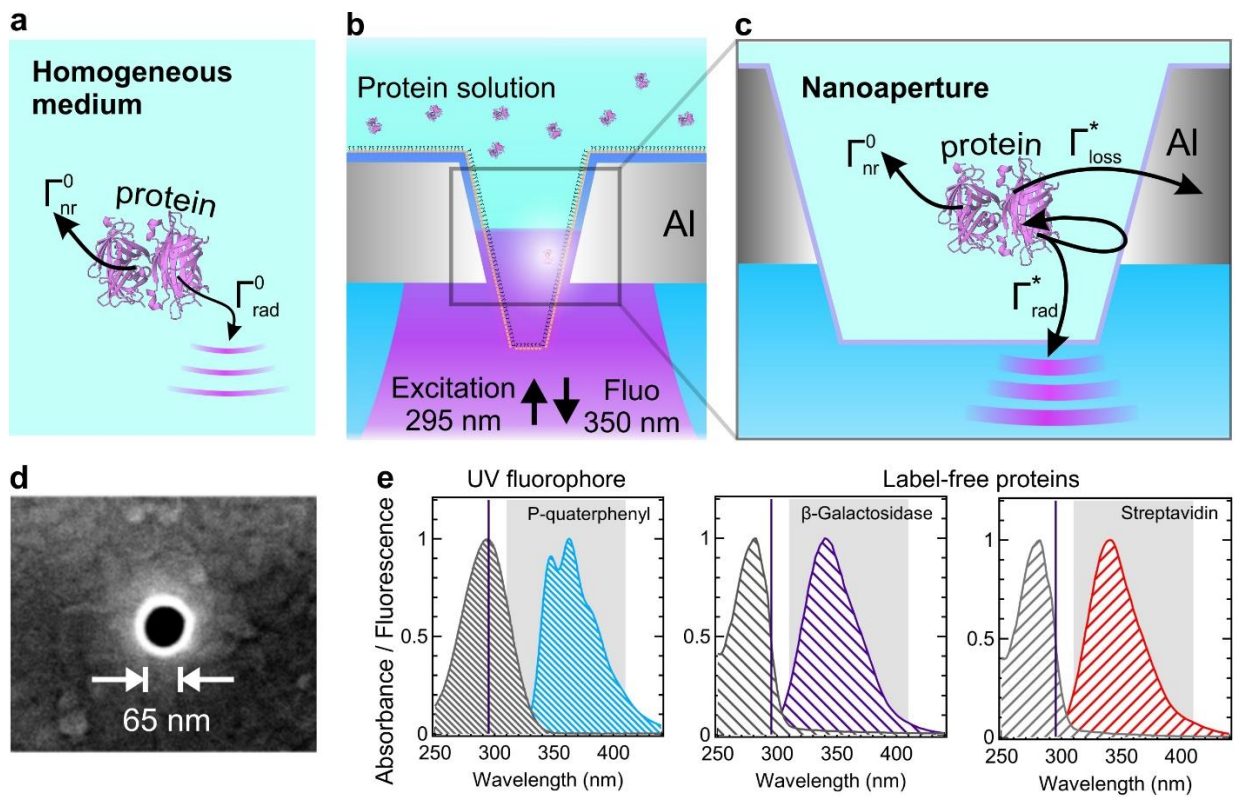

**Figure 1.** Aluminum nanoaperture to enhance the radiative rate of label-free proteins in the UV. (a) Notations describing the decay pathways of protein excited in the homogeneous space (free solution). (b) Schematic view of the experiment in the presence of Al nanoapertures. (c) Notations describing the protein decay pathways modified in the presence of nanoapertures. (d) Scanning electron microscope image of a 65 nm diameter nanoaperture. (e) Normalized absorbance and fluorescence spectra of the UV dye p-quaterphenyl and two proteins β-galactosidase and streptavidin. The violet vertical lines correspond to the excitation wavelength of 295 nm, while the shaded regions show the wavelength range used for fluorescence detection.

**Methods**

*Nanoaperture fabrication.* A 100 nm-thick layer of aluminum was deposited on top of cleaned quartz coverslips using electron-beam evaporation (Bühler Syrus Pro 710). The deposition parameters (chamber pressure $10^{-6}$ mbar, deposition rate 10 nm/s) were set to reduce the amount of aluminum oxide in the bulk metal [27]. Nanoapertures with 65 nm diameter (Figure 1d) were milled using gallium-based focused ion beam (FEI dual beam DB235 Strata, voltage 30 kV, ion current 10 pA). This nanoaperture diameter corresponds to a cut-off diameter in the UV [15], where the real part of the propagation constant vanishes and the intensity evanescently decays inside the nanoaperture. While smaller diameters would further promote the localization of electric field and the excitation intensity enhancement, significant quenching losses occur for diameter below the cut-off, which decrease the overall fluorescence enhancement [14]. The 65 nm diameter used here provides the highest fluorescence enhancement for the nanoaperture. The fabrication includes a 50 nm undercut into the quartz substrate to optimize the fluorescence enhancement [13,15,32].

*Nanoaperture protection.* To protect the aluminum surface against the accelerated photocorrosion in the UV [33], we deposited a 10 nm-thick $SiO_2$ layer by plasma-enhanced chemical vapor protection (PECVD) using a PlasmaPro NGP80 from Oxford Instruments. After a pumping step of 20 minutes at 300 °C, the $SiO_2$ layer is deposited at the same temperature using two precursors: 5% $SiH_4/N_2$ and $N_2O$ with flows of 160 and 710 sccm, respectively. The chamber pressure is set at 1000 mTorr and high-frequency power of 20 W. The silica is deposited at the rate of 1 nm/s.

*Nanoaperture passivation.* To avoid the non-specific adhesion of β-galactosidase and streptavidin proteins on the surfaces, we passivated the samples with polyethylene glycol (PEG)[34]. Silane-modified polyethylene glycol (PEG) of molecular weight 1000 Da (PEG 1000) is purchased from NANOCS. The PEG-silane is dissolved at the concentration of 1 mg/ml in ethanol with 1 v/v% acetic acid. The solution of the PEG-silane is placed on the surface of the nanoapertures covered with the $SiO_2$ layer for 3-4 hours, resulting in a satisfactory monolayer PEG-coating of the nanoaperture interior surface.

*UV dye and protein samples.* P-quaterphenyl is a UV fluorescent dye featuring a high quantum yield of 89 %, which serves here as a reference to calibrate the nanoaperture influence in the UV. P-quaterphenyl (purity >98%) was purchased from TCI and dissolved in cyclohexane. As natural (label-free) proteins, we have selected β-galactosidase and streptavidin which contain respectively 156 and 24 tryptophan amino-acid residues in their sequence. Tryptophan emission in proteins is often inherently quenched due to electron transfers to the neighboring amino acids or energy homotransfer among the tryptophan residues in the protein structure [18,35]. Thus, β-galactosidase and streptavidin possess low quantum yields of 1.6 and 3.5 %, respectively [36].

β-galactosidase from *Escherichia coli* (156 tryptophan residues, 466 kDa, Sigma-Aldrich) was stored in Hepes buffer (Hepes 25 mM, NaCl 300 mM, 0.1 v/v% Tween20, DTT 1 mM, EDTA 1mM, pH 7) at −20 °C temperature. For the measurements, the stock solution was diluted in a Hepes buffer including the GODCAT oxygen scavenger [37] (100 nM glucose oxidase, 830 nM catalase, 10 w/v% D-glucose), 0.5% Tween20, and 2 mM urea at pH 7. GODCAT was added to deplete the oxygen dissolved in the buffers and improve the UV photostability [36–38].

Streptavidin from *Streptomyces avidinii* (24 tryptophan residues, 60 kDa, Sigma-Aldrich) was stored in a Hepes buffer (Hepes 25 mM, NaCl 100 mM, 0.1 v/v% Tween20). For the measurements, the stock solution was diluted in a Hepes buffer including GODCAT and 10 mM ascorbic acid at pH 4 to promote photocorrosion resistance [33]. Figure 1e shows the absorption and fluorescence spectra of p-quaterphenyl, β-galactosidase, and streptavidin recorded using a cuvette spectrophotometer (Tecan Spark 10M).

*Experimental setup.* The experiments were performed on a custom-built confocal microscope described in detail in ref.[14] using a 295 nm picosecond laser (Picoquant VisUV-295-590, 70 ps pulse duration, 80 MHz repetition rate). With this 295 nm excitation, only the tryptophan residues contributed to the detected protein intrinsic emission, as the other aromatic amino acids such as tyrosine and phenylalanine are not excited above 290 nm [18]. The experiments on p-quaterphenyl were implemented using a Zeiss Ultrafluar 40x, 0.6 NA, glycerol immersion objective in the epi-fluorescence configuration at the laser power of 120 µW. The proteins were measured with a LOMO 58x, 0.8 NA, water immersion objective. β-galactosidase and streptavidin were excited at 10 µW and 50 µW, respectively. A dichroic mirror (Semrock FF310-Di01-25-D) and two emission filters (Semrock FF01-300/LP-25 and Semrock FF01-375/110-25) filtered the fluorescence light in the wavelength range between 310 and 410 nm, which covered a major part of the fluorescence spectra from the samples (Fig. 1e). The detection was performed by a single photon-counting photomultiplier tube (Picoquant PMA 175) after a 50 µm pinhole conjugated to the object plane. The photomultiplier tube output was recorded with a time-correlated single photon counting module (Picoharp 300, Picoquant) featuring 150 ps overall temporal resolution. Correlation functions in the nanoapertures were obtained for both β-galactosidase and streptavidin proteins within 2 min acquisition.

*FCS analysis.* Fluorescence correlation spectroscopy (FCS) is used to analyze the temporal fluctuations of the fluorescence intensity on a microsecond to second timescale [39].

FCS involves computing the intensity-normalized correlation function $G(t) = \langle\delta I(0)\delta I(t)\rangle/\langle I(t)\rangle^2$, where $\delta I(t) = I(t) - \langle I\rangle$ is the intensity fluctuation around the average. From this function, we extract the average number of detected molecules allowing to compute the average fluorescence brightness per molecule without any assumption on the molecular concentration or the size of the detection volume. The FCS correlation data was fitted by Levenberg-Marquardt optimization using the commercial software SymPhoTime 64 (Picoquant). The three-dimensional Brownian diffusion model was employed to describe the data as in the previous works [14,15].

$$G(\tau) = \frac{1}{N_{mol}}\left[1 - \frac{B}{F}\right]^2 \left(1 + \frac{\tau}{\tau_d}\right)^{-1} \left(1 + \frac{1}{\kappa^2}\frac{\tau}{\tau_d}\right)^{-0.5} \qquad (1)$$

where $N_{mol}$ denotes the number of molecules in the detection volume, $B$ is the background noise intensity from the photostabilizing buffer and the empty nanoaperture itself, $F$ is referred to the total fluorescence intensity, $\tau_d$ denotes the mean diffusion time, and $\kappa$ is the aspect ratio of the axial to lateral dimensions of the detection volumes ($\kappa = 8$ for the confocal illumination and $\kappa = 1$ for the nanoaperture). This model was found to sufficiently describe the FCS correlation data [14,15,30]. As we are primarily interested here in the number of molecules and their fluorescence brightness, the choice of the FCS model for the translational diffusion plays no specific role. All the fit results from the FCS analysis for the different cases are summarized in Table 1.

*Fluorescence lifetime analysis.* The fluorescence decay histograms were fitted using Picoquant SymPhoTime 64 software with an iterative reconvolution fit that considered the instrument response function (IRF). The fits were performed in the data range accounting for 95% of the total collected photons. For the confocal reference for p-quaterphenyl, a single-exponential fit was enough to interpolate the lifetime decay owing to the large fluorescence signal and the low noise from the cyclohexane solvent. For all the other cases, we detected background noise in the UV, stemming from fast-lived photoluminescence of aluminum, gallium ions, organic additives, and laser back-scattered light, we fixed one component $\tau_1 = 0.01$ ns for the fits and this contribution was not considered for the determination of fluorescence lifetime from the sample. For the fluorescence decay of p-quaterphenyl in nanoapertures, we observed that in order to take well into account the decay tail at long time delays, we had to consider a contribution with a fixed lifetime of 0.76 ns corresponding to the lifetime of p-quaterphenyl in the confocal case. This was referred to as a residual fluorescence contribution from molecules lying away from the aperture whose fluorescence emission was not enhanced by the plasmonic nanostructure. Therefore, the read-out lifetime value of p-quaterphenyl corresponds to the second $\tau_2$ component. We used the same procedure for the fluorescence decays of β-galactosidase and streptavidin proteins in nanoapertures. The fluorescence decays in the free solution yielded two exponential components which stem from the formation of rotational isomers of tryptophan [40]. Thereby, the read-out lifetimes for the proteins in the free solution and the nanoaperture were referred to as the intensity average of the two components $\tau_2$ and $\tau_3$. All the fit results for the lifetime analysis are summarized in Table 2.

**Results**

Figure 1a-c introduces the concept of the protein UV emission modification in the presence of Al nanoapertures and our notations. In the homogeneous solution case, the protein total decay rate constant $\Gamma_{tot}^0$ is the sum of the radiative rate constant $\Gamma_{rad}^0$ and the internal non-radiative decay rate constant $\Gamma_{nr}^0$. In the case of protein UV autofluorescence, the internal non-radiative rate $\Gamma_{nr}^0$ generally largely exceeds the radiative rate $\Gamma_{rad}^0$ so that the quantum yield $\varphi = \Gamma_{rad}^0/(\Gamma_{rad}^0 + \Gamma_{nr}^0)$ is around 1 to 10% depending on the protein of interest [41]. The presence of the plasmonic nanoaperture modifies the local density of optical states (LDOS) [4,20] and accelerates the relaxation rate from the excited state to the ground state. The radiative rate $\Gamma_{rad}^*$ is enhanced, and we define the Purcell factor $\eta_{\Gamma_{rad}} = \Gamma_{rad}^*/\Gamma_{rad}^0$ as the enhancement of the radiative decay rate introduced by the plasmonic nanostructure. The occurrence of quenching losses into the metal opens an additional energy transfer pathway with a rate constant $\Gamma_{loss}^*$. The total decay rate constant becomes $\Gamma_{tot}^* = \Gamma_{rad}^* + \Gamma_{nr}^0 + \Gamma_{loss}^*$ as we assume that the internal conversion rate $\Gamma_{nr}^0$ is unchanged by the photonic environment (losses introduced by the photonic environment are accounted by $\Gamma_{loss}^*$). In the following, we aim at measuring experimentally these different decay rate constants for two different proteins (β-galactosidase and streptavidin) to quantify the influence of a plasmonic nanoaperture on the LDOS and explore if Purcell enhancement can be demonstrated for label-free proteins. We also compare our results with p-quaterphenyl, a well-established high quantum yield fluorescent dye in the UV range [42]. While it is now well established that a single fluorescent dye can be approximated by a point electromagnetic dipole whose radiation can be controlled by the LDOS [1], the extension of the LDOS enhancement and Purcell effect to protein UV autofluorescence is not completely straightforward. Each protein contains several tens of tryptophan emitters which are closely packed at a nanometer distance and are surrounded by other amino acids acting as fluorescence quenchers *via* electron or proton transfer [17].

Figure 2 gathers the raw experimental data for all the three investigated molecules. Each line of three graphs corresponds to a different molecule and shows the fluorescence intensity time traces, the FCS correlation functions computed from the intensity time traces, and the corresponding fluorescence lifetime decays, for both the 65 nm diameter nanoaperture and the reference confocal case. All the fit results for the different cases are summarized in Table 1 for the FCS analysis and Table 2 for the lifetime analysis.

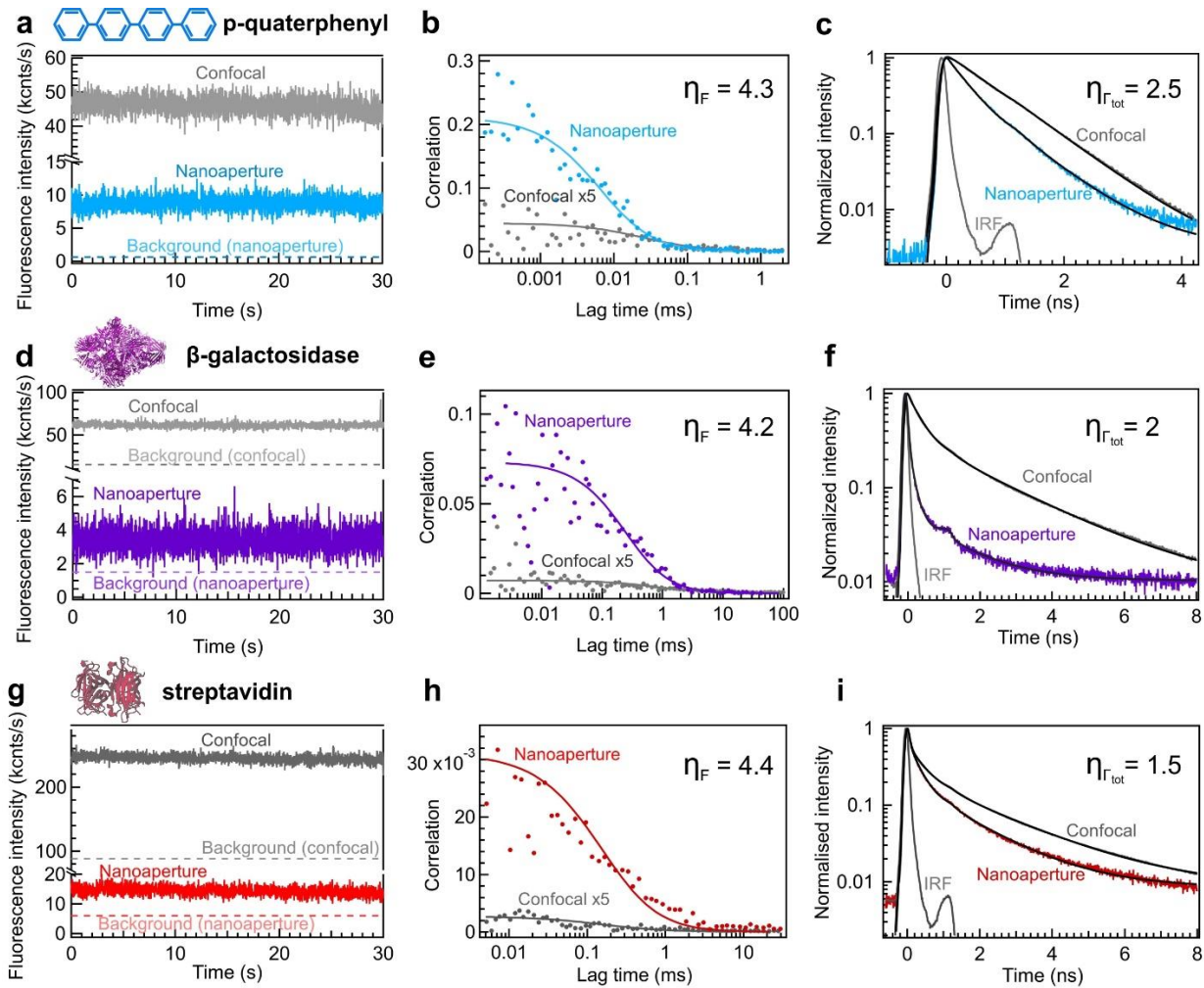

**Figure 2.** Experimental characterization of intrinsic protein fluorescence and total decay rate in the Al nanoaperture of 65 nm diameter. (a) Fluorescence intensity time traces, (b) FCS correlation functions, and (c) fluorescence lifetime decay of diffusing p-quaterphenyl molecules under confocal illumination and in the nanoaperture. P-quaterphenyl concentration in cyclohexane is 20 nM in the free solution and 2 μM in the nanoaperture. (d) Fluorescence intensity time traces, (e) FCS correlation functions, and (f) fluorescence lifetime decay of β-galactosidase protein for the confocal reference and the nanoaperture. β-galactosidase concentration in the aqueous buffer is 0.5 μM in the free solution and 1.7 µM in the nanoaperture. (g) Fluorescence intensity time traces, (h) FCS correlation functions, and (i) fluorescence lifetime decay of streptavidin protein for the confocal reference and the nanoaperture. Streptavidin concentration in the aqueous buffer is 1.5 µM in the free solution and 6 µM in the nanoaperture. The values of fluorescence enhancement corresponding to each UV emitting molecule are shown in panels (b,e,h), while the gains of total decay rate are indicated in panels (c,f,i). IRF denotes the instrument response function of the experimental setup.

**Table 1.** FCS fit results corresponding to the correlation data displayed in Figure 2.

|  | $F$ (kHz) | $B$ (kHz) | $N_{mol}$ | $\tau_d$ (µs) | CRM (kHz) |
|---|---|---|---|---|---|
| P-quaterphenyl (confocal) | 45 | 0 | 112 | 25 | 0.4 |
| P-quaterphenyl (nanoaperture) | 7.7 | 0.65 | 4 | 10 | 1.79 |
| β-galactosidase (confocal) | 57.9 | 15.5 | 380 | 800 | 0.11 |
| β-galactosidase (nanoaperture) | 3.6 | 1.5 | 4.6 | 367 | 0.44 |
| Streptavidin (confocal) | 240 | 90 | 870 | 220 | 0.17 |
| Streptavidin (nanoaperture) | 13.4 | 5.9 | 10.1 | 230 | 0.75 |

**Table 2.** Fit results for the fluorescence lifetime analysis of the histograms depicted in Figure 2.

|  | $\tau_1$ (ns) | $\tau_2$ (ns) | $\tau_3$ (ns) | $I_1$ | $I_2$ | $I_3$ | $\tau_{tot}$ |
|---|---|---|---|---|---|---|---|
| P-quaterphenyl (confocal) | - | 0.76 | - | - | 1 | - | 0.76 |
| P-quaterphenyl (nanoaperture) | 0.01 | 0.31 | 0.76 | 0.13 | 0.38 | 0.49 | 0.31 |
| β-galactosidase (confocal) | 0.01 | 0.35 | 2.17 | 0.08 | 0.25 | 0.67 | 1.68 |
| β-galactosidase (nanoaperture) | 0.01 | 0.14 | 1.32 | 0.64 | 0.15 | 0.21 | 0.85 |
| Streptavidin (confocal) | 0.01 | 0.39 | 2.01 | 0.2 | 0.2 | 0.6 | 1.61 |
| Streptavidin (nanoaperture) | 0.01 | 0.28 | 1.48 | 0.27 | 0.25 | 0.48 | 1.07 |

The FCS data on proteins exhibit a higher level of noise as compared to p-quaterphenyl. The FCS signal-to-noise ratio depends linearly on the fluorescence brightness per molecule (CRM) [43], which is lower for the proteins as compared to p-quaterphenyl. This highlights an additional advantage of the nanoaperture for UV-FCS: the higher photon count rates reduce the statistical noise and allow decreasing the experiment integration time. These features are important as in the UV, the microscope objectives have a limited numerical aperture for fluorescence collection, and the target biomolecules have a quite low photostability and quantum yield as compared to conventional fluorescent dyes in the visible [36].

From the fit of the FCS data, we measure the average number of molecules $N_{mol}$ detected for each experiment, which allows us to compute the fluorescence brightness per molecule as the ratio of the average total fluorescence intensity (corrected for the background) by the average number of fluorescent molecules CRM = F-B / $N_{mol}$. Please note that this molecular brightness is averaged over all the spatial positions and orientations of the molecules inside the detection volume. We then determine the net fluorescence enhancement $\eta_F$ as the CRM increase in the presence of the nanoaperture. For p-quaterphenyl, the molecular brightness increases from 0.4 kcounts/s to 1.8 kcounts/s in presence of the nanoaperture, corresponding to a fluorescence enhancement $\eta_F$ of 4.3 ± 0.3. Similar fluorescence enhancement values are found with β-galactosidase and streptavidin proteins, with 4.2 ± 0.4 and 4.4 ± 0.4 respectively.

From the fit of the fluorescence decay histograms, we determine the average fluorescence lifetime $\tau$, which corresponds to the inverse of the total decay rate constant $\Gamma_{tot} = 1/\tau$. The reduction of the fluorescence lifetime in the presence of the nanoaperture quantifies the gain of the total decay rate $\eta_{\Gamma_{tot}} = \Gamma^*_{tot}/\Gamma^0_{tot}$. For p-quaterphenyl, the fluorescence lifetime is reduced from 0.76 ns to 0.3 ns with the nanoaperture, so that the total decay rate is enhanced by $\eta_{\Gamma_{tot}}$ = 2.5 ± 0.2. The fluorescence lifetimes measured for proteins are also shortened in the presence of the nanoaperture, with an acceleration of $\eta_{\Gamma_{tot}}$ = 2.0 ± 0.3 for β-galactosidase and 1.5 ± 0.2 for streptavidin. However, while the shorter autofluorescence lifetime for proteins in the nanoaperture indicates a change in the LDOS, it does not constitute by itself a demonstration of the radiative rate enhancement. Indeed, the presence of quenching losses to the metal $\Gamma^*_{loss}$ contribute to the observation of an increased total decay rate in presence of the nanoaperture, and this contribution must be taken into account. To disentangle the modification of the radiative rate from the total decay rate, we analyze further the fluorescence brightness enhancement.

The net fluorescence enhancement η$_F$ in the vicinity of the nanoaperture depends on three factors: the increase of the excitation intensity, quantum yield (φ), and the collection efficiency [30]. It can be expressed as:

$$\eta_F = \frac{\eta_{coll}\, \eta_{exc}\, \eta_{\Gamma_{rad}}}{\eta_{\Gamma_{tot}}} \quad (2)$$

where $\eta_{coll}$ is the gain in collection efficiency, $\eta_{exc}$ is the local excitation intensity enhancement and $\eta_{\Gamma_{rad}}/\eta_{\Gamma_{tot}}$ corresponds to the quantum yield gain. Our experiments in Fig. 2 measure the gains $\eta_F$ and $\eta_{\Gamma_{tot}}$. To extract the radiative decay rate enhancement $\eta_{\Gamma_{rad}}$ (the Purcell factor), we use two extra pieces of information. First, the antenna reciprocity theorem states the excitation intensity gain $\eta_{exc}$ amounts to the product of the collection efficiency gain $\eta_{coll}$ and the radiative rate gain $\eta_{\Gamma_{rad}}$ [4,44]. Second, we set the collection efficiency gain $\eta_{coll}$ = 1.44 for the 65 nm nanoaperture, relying on the previous descriptions of the directionality of bare aluminum nanoapertures [45,46]. Altogether, the radiative rate enhancement becomes $\eta_{\Gamma_{rad}} = \sqrt{\eta_F\, \eta_{\Gamma_{tot}}}/1.44$, and we use this expression for all the three different fluorescent samples (proteins and p-quaterphenyl). Moreover, once all the rate enhancements are known, then the knowledge of the reference fluorescence quantum yield in the confocal case allows us to compute back all the different decay rate constants contributing to the fluorescence process both in the confocal and nanoaperture case.

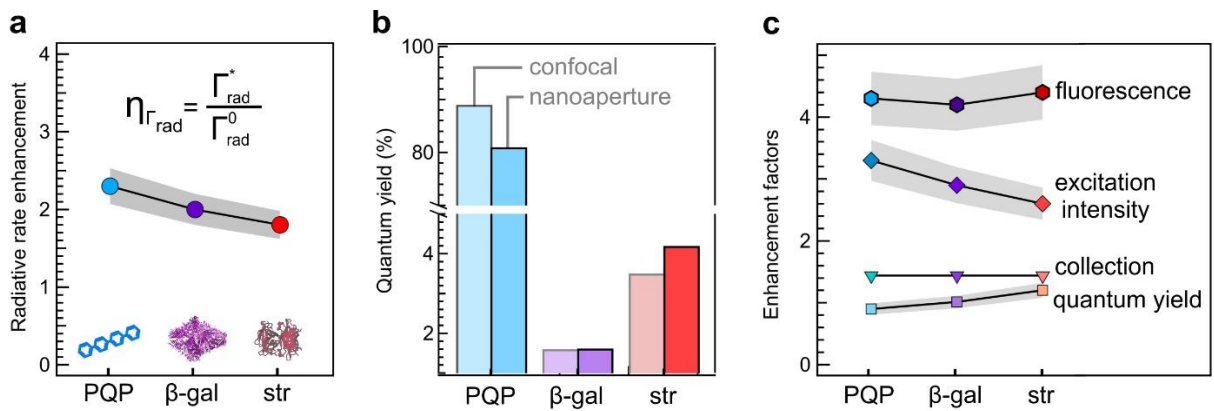

**Figure 3.** Enhancement of the different decay rate constants and comparison between label-free proteins and p-quaterphenyl (PQP). (a) Radiative rate enhancement (Purcell factor). (b) The average quantum yield of the UV fluorescent molecules. The data for the confocal reference is displayed in pastel colors, while the brighter colors represent the nanoaperture. (c) Excitation intensity gain, collection efficiency gain, quantum yield gain, and net fluorescence enhancement. The gray shaded areas represent the error bars.

**Table 3.** Photokinetic rate constants deduced from the experimental data. All $\Gamma_i$ values are expressed in ns$^{-1}$.

| Sample | Configuration | $\Gamma_{rad}$ | $\Gamma_{nr}$ | $\Gamma_{loss}$ | $\Gamma_{tot}$ | $\tau_{tot}$ (ns) | $\varphi$ (%) |
|---|---|---|---|---|---|---|---|
| p-quaterphenyl | Confocal | 1.17 | 0.15 | - | 1.32 | 0.76 | 89 |
| | Nanoaperture | 2.68 | 0.15 | 0.5 | 3.33 | 0.3 | 81 |
| β-galactosidase | Confocal | 0.01 | 0.59 | - | 0.6 | 1.68 | 1.6 |
| | Nanoaperture | 0.02 | 0.59 | 0.57 | 1.18 | 0.85 | 1.6 |
| Streptavidin | Confocal | 0.02 | 0.6 | - | 0.62 | 1.61 | 3.5 |
| | Nanoaperture | 0.04 | 0.6 | 0.29 | 0.93 | 1.07 | 4.2 |

Figure 3 and Table 3 summarize our main results regarding the UV fluorescence photokinetic rates enhancement for label-free proteins inside an aluminum nanoaperture. We measure a radiative rate enhancement (Purcell factor) of 2.0 ±0.2 x for β-galactosidase and 1.8 ± 0.2 x for streptavidin. These values correspond also to the radiative rate enhancement of 2.3 ± 0.2 x for found for p-quaterphenyl (Fig. 3a), which is within the value range computed for Al nanoapertures of similar design independently [47]. Remarkably, these results are achieved even though the fluorescence quantum yield between p-quaterphenyl and the proteins differ up to nearly two orders of magnitude (Fig. 3b). Altogether, this provides the first clear demonstration of radiative enhancement and Purcell effect applied on label-free proteins in the UV. These observations validate the possibility to control protein intrinsic autofluorescence by modifying the LDOS with plasmonic and nanophotonic devices. Importantly, in our approach, the fluorescence enhancement is measured based on FCS calibration of the number of molecules, implying that the resulting gains are not biased by possible concentration variations of the proteins.

Figure 3c shows all the deduced enhancement factors that constitute the net fluorescence gain measured in the UV. Despite the very different nature of the fluorescent emitters, the parameters retrieved for the influence of the photonic structure are similar, confirming the validity of our approach. Moreover, the derived excitation intensity gain goes along with the simulations reported in the previous work [14].

Based on the deduced enhancement factors, we can reveal the decay rates of $\Gamma^*_{rad}$ and $\Gamma^*_{loss}$. For p-quaterphenyl, which is a high quantum yield dye, the radiative rate constant $\Gamma^0_{rad}$ dominates over the internal non-radiative rate $\Gamma^0_{nr}$. Therefore, the Purcell enhancement of the radiative rate does not affect substantially the quantum yield. The losses $\Gamma^*_{loss}$ induced by coupling to the metal induce a slight reduction of its quantum yield from 89 to 81 % inside the nanoaperture. On the contrary, proteins are intrinsically poor emitters, their radiative rate constants $\Gamma^0_{rad}$ are small as compared to their internal non-radiative rate $\Gamma^0_{nr}$ (Fig. 3b, Tab. 3). In this case, the quantum yields are increased inside the nanoaperture thanks to the radiative rate enhancement, with a factor of 1.2× for streptavidin (Fig. 3b,c). Lastly, we point out that the quenching losses induced by the metal structure $\Gamma^*_{loss}$ remain of similar amplitude for all three investigated molecules.

**Conclusion**

We report the observation and thorough quantification of the Purcell effect for the intrinsic UV emission of label-free β-galactosidase and streptavidin proteins in plasmonic nanoapertures. The detected spontaneous emission originates from the constituent tryptophan residues, implying that our conclusion and the techniques described here are valid for most of the proteins. Regardless of the complexity of protein structure and its dominating non-radiative decay rate, we measure a spontaneous emission gain matching that of a high quantum yield p-quaterphenyl UV fluorescent dye. This shows that the concepts and results of nanophotonics to control the LDOS can also be largely applied to label-free proteins featuring hundreds or thoursands of aminoacids. Our observations importantly show ways to improve the net detected autofluorescence signal by increasing the radiative emission and the autofluorescence quantum yield. Altogether, our experiments demonstrate the possibility to control and measure the intrinsic spontaneous emission of single protein molecules in the UV, which is an important finding for future development of label-free protein detection and sensing.

**Notes**

The authors declare no competing financial interest.**Acknowledgments**

The authors thank Antonin Moreau and Julien Lumeau for help with the aluminum deposition and Marco Abbarchi for help with the PECVD. This project has received funding from the European Research Council (ERC) under the European Union's Horizon 2020 research and innovation programme (grant agreement No 723241).
**References**

[1]  Barnes W L, Horsley S A and Vos W L 2020 Classical antennas, quantum emitters, and densities of optical states *J. Opt.* **22** 073501

[2]  Lodahl P, Driel A F van, Nikolaev I S, Irman A, Overgaag K, Vanmaekelbergh D and Vos W L 2004 Controlling the dynamics of spontaneous emission from quantum dots by photonic crystals *Nature* **430** 654–7

[3]  Somaschi N, Giesz V, De Santis L, Loredo J C, Almeida M P, Hornecker G, Portalupi S L, Grange T, Antón C, Demory J, Gómez C, Sagnes I, Lanzillotti-Kimura N D, Lemaítre A, Auffeves A, White A G, Lanco L and Senellart P 2016 Near-optimal single-photon sources in the solid state *Nat. Photonics* **10** 340–5

[4]  Novotny L and Van Hulst N 2011 Antennas for light *Nat. Photonics* **5** 83–90

[5]  Koenderink A F 2010 On the use of Purcell factors for plasmon antennas *Opt. Lett.* **35** 4208–10

[6]  Akselrod G M, Argyropoulos C, Hoang T B, Ciracì C, Fang C, Huang J, Smith D R and Mikkelsen M H 2014 Probing the mechanisms of large Purcell enhancement in plasmonic nanoantennas *Nat. Photonics* **8** 835–40

[7]  Anger P, Bharadwaj P and Novotny L 2006 Enhancement and quenching of single-molecule fluorescence *Phys. Rev. Lett.* **96** 113002

[8]  Khatua S, Paulo P M R, Yuan H, Gupta A, Zijlstra P and Orrit M 2014 Resonant Plasmonic Enhancement of Single-Molecule Fluorescence by Individual Gold Nanorods *ACS Nano* **8** 4440–9

[9]  Frimmer M, Chen Y and Koenderink A F 2011 Scanning emitter lifetime imaging microscopy for spontaneous emission control *Phys. Rev. Lett.* **107** 123602
14